\documentclass{article}
\usepackage{fullpage}
\usepackage{amssymb}
\usepackage{amsmath}
\usepackage{graphicx}
\usepackage{algorithm}
\usepackage[noend]{algorithmic}
\renewcommand{\algorithmicrequire}{\textbf{Input}}
\renewcommand{\algorithmicensure}{\textbf{Output}}
\usepackage{microtype}
\usepackage{color}
\usepackage{hyperref}

\renewcommand{\O}{O}

\def\G{\mathcal{G}}

\def\leq{\leqslant}
\def\geq{\geqslant}
\newcommand{\Reals}{\mathbb{R}}
\newcommand{\dil}{\delta}
\newcommand{\Dil}{\Delta}
\newcommand{\dilConv}{\dil_{\mathrm{C}}}

\newcommand{\eps}{\varepsilon}

\newtheorem{theorem}{Theorem}
\newtheorem{lemma}{Lemma}

\newtheorem{corollary}{Corollary}

\makeatletter
\newbox\ProofSym
\setbox\ProofSym=\hbox{%
\unitlength=0.18ex%
\begin{picture}(10,10)
\put(0,0){\framebox(9,9){}}
\put(0,3){\framebox(6,6){}}
\end{picture}}
\newenvironment{proof}[1][Proof.]{\O@proof{#1}}{\O@endproof}
\def\O@proof#1{\trivlist
  \@topsep\z@\@topsepadd\smallskipamount%
  \@ifstar{\item[]}{\item[\hskip\labelsep\it #1 ]}}
\def\O@endproof{\endtrivlist}
\def\qed{\hfill\copy\ProofSym}

\let\geq\geqslant
\let\leq\leqslant

\makeatletter
\partopsep\z@ \textfloatsep 10pt plus 1pt minus 4pt
\def\section{\@startsection {section}{1}{\z@}{-3.5ex plus -1ex minus
-.2ex}{2.3ex plus .2ex}{\large\bf}}
\def\subsection{\@startsection{subsection}{2}{\z@}{-3.25ex plus -1ex
minus -.2ex}{1.5ex plus .2ex}{\normalsize\bf}}
\def\@fnsymbol#1{\ensuremath{\ifcase#1\or *\or 1\or 2\or 3\or 4\or 5\or
    6\or 7\or 8\or 9\else\@ctrerr\fi}}
\makeatother

\title{Sparse geometric graphs with small dilation%
  \thanks{BA was supported in part by NSF ITR Grant CCR-00-81964
    and by a grant from the US-Israel Binational Science Foundation.  Part of
    the work was carried out while BA was visiting TU/e in February 2004 and
    in the summer of 2005.
    OC was supported by LG Electronics.
    MdB was supported by the Netherlands' Organisation for Scientific
    Research (NWO) under project no.~639.023.301.
    MS was supported by the Natural Sciences and Engineering Research
    Council of Canada (NSERC).
    AV was supported by NUS research grant R-252-000-166-112.}}

\author{Boris~Aronov%
  \thanks{Department of Computer and
    Information Science, Polytechnic University, Brooklyn, New York, USA.
    {http://cis.poly.edu/\~{}aronov}}
  \and
  Mark~de~Berg%
  \thanks{Department of Mathematics and Computing Science, TU Eindhoven,
    Eindhoven, the Netherlands.
    {mdberg@win.tue.nl},
    {cs.herman@haverkort.net}}
  \and
  Otfried~Cheong%
  \thanks{Division of Computer Science, KAIST, Daejeon, South
    Korea. {otfried@tclab.kaist.ac.kr}}
  \and
  Joachim~Gudmundsson%
  \thanks{National ICT Australia Ltd,
    Australia. {joachim.gudmundsson@nicta.com.au}.
    NICTA is funded through the Australian
    Government's Backing Australia's Ability initiative, in part through
    the Australian Research Council.}
  \and
  Herman~Haverkort%
  \footnotemark[3]
  \and
  Michiel~Smid%
  \thanks{School of Computer Science, Carleton University, Ottawa,
    Canada. {michiel@scs.carleton.ca}}
  \and
  Antoine~Vigneron%
  \thanks{Unit\'e Math\'ematiques et Informatique Appliqu\'ees,
    INRA, Jouy-en-Josas, France.
    {antoine.vigneron@jouy.inra.fr}}}

\begin{document}
\maketitle

\begin{abstract}
  Given a set $S$ of $n$ points in $\Reals^D$, and an integer $k$ such
  that $0\leq k< n$, we show that a geometric graph with vertex
  set~$S$, at most $n - 1 + k$ edges, maximum degree five, and
  dilation $O(n / (k+1))$ can be computed in time $O(n \log n)$. For
  any $k$, we also construct planar $n$-point sets for which any
  geometric graph with $n-1+k$ edges has dilation $\Omega(n/(k+1))$; a
  slightly weaker statement holds if the points of $S$ are required to
  be in convex position.
\end{abstract}

\section{Preliminaries and introduction}

A \emph{geometric network} is an undirected graph whose vertices are points
in $\Reals^D$. Geometric networks, especially geometric networks of points
in the plane, arise in many applications. Road networks, railway networks,
computer networks---any collection of objects that have some connections
between them can be modeled as a geometric network.
A natural and widely studied type of geometric network is the
\emph{Euclidean network}, where the weight of an edge is simply
the Euclidean distance between its two endpoints.
Such networks for points in $\Reals^D$ form the topic of study
of our paper.

When designing a network for a given set $S$ of points, several
criteria have to be taken into account.  In particular, in many
applications it is important to ensure a short connection between
every two points in $S$. For this it would be ideal to have a direct
connection between every two points; the network would then be a
complete graph. In most applications, however, this is unacceptable
due to the high costs. Thus the question arises: is it possible to
construct a network that guarantees a reasonably short connection
between every two points while not using too many edges? This leads to
the well-studied concept of \emph{spanners}, which we define next.

The weight of an edge $e=(u,v)$ in a Euclidean network $G=(S,E)$ on a
set $S$ of $n$ points is the Euclidean distance between $u$ and $v$,
which we denote by $d(u,v)$.  The \emph{graph distance} $d_G(u,v)$
between two vertices $u,v\in S$ is the length of a shortest path in
$G$ connecting $u$ to $v$.  The \emph{dilation} (or \emph{stretch
factor}) of $G$, denoted $\Dil(G)$, is the maximum factor by which the
graph distance $d_G$ differs from the Euclidean distance $d$, namely
\[
\Dil(G):= \max_{\substack{u,v \in S \\ u \neq v}} \frac{d_G(u,v)}{d(u,v)}.
\]
The network $G$ is a \emph{$t$-spanner} for $S$ if $\Dil(G) \leq t$.

Spanners find applications in robotics, network topology
design, distributed systems, design of parallel machines, and many
other areas and have been a subject of considerable
research~\cite{cdns-nsrgs-95}. Recently spanners found
interesting practical applications in metric space searching
\cite{np-pcmts-03,npc-tsdsm-02} and broadcasting in communication
networks \cite{alwwf-gswah-03,fpzw-smdn-04,l-acgwa-03}. The problem of
constructing spanners has received considerable attention from both an
experimental perspective~\cite{fg-esgs-05,sz-cmws-04} and theoretical
perspective---see the surveys~\cite{e-sts-00,gk-ddgn-07,s-cppcg-00} 
and the book by Narasimhan and Smid~\cite{ns-gsn-07}.

The complete graph has dilation~1, which is optimal, but
we already noted that the complete graph is generally too costly.
The main challenge is therefore to design \emph{sparse}
networks that have small dilation.
There are several possible measures of sparseness, for example
the total weight of the edges or the maximum degree of a vertex.
The measure that we will focus on is the number of edges.
Thus the main question we study is this:
Given a set $S$ of $n$ points in $\Reals^D$,
what is the best dilation one can achieve with a network
on $S$ that has few edges? Notice that the edges
of the network are allowed to cross or overlap.

This question has already received ample attention.
For example, there are several
algorithms~\cite{ck-dmpsa-95,ll-tapga-92,s-cmsg-91,v-sgagc-91}
that compute a $(1+\eps)$-spanner for $S$, for any given constant
$\eps>0$. The number of edges in these spanners is $O(n)$.
Although the number of edges is linear in $n$, it can still be rather
large due to the hidden constants in the $O$-notation that depend
on $\eps$ and the dimension $D$.
Das and Heffernan~\cite{dh-cdsosp-96} showed how to compute in
$O(n \log n)$ time, for any constant $\eps'>0$, a $t$-spanner with
$(1+\eps')n$ edges and degree at most three, where $t$ only depends on
$\eps'$ and $D$.

Any spanner must have at least $n-1$ edges, for otherwise the graph
would not be connected, and the dilation would be infinite.  In this
paper, we are interested in the case when the number of edges is close
to~$n-1$, not just linear in~$n$.  This leads us to define the
quantity $\Dil(S,k)$:
\[
\Dil(S,k) := \min_{\substack{V(G)=S\\|E(G)|=n-1+k}} \Dil(G).
\]
Thus $\Dil(S,k)$ is the minimum dilation one can achieve with a
network on $S$ that has $n-1+k$ edges. 



Klein and Kutz~\cite{kk-cgmdg-06} showed recently that given a set of
points $S$, a dilation $t$ and a number $k>0$, it is NP-hard to decide
whether $\Dil(S, k) \leq t$.  Cheong et al.~\cite{chl-cmdst-07} showed
that even the special case $k=0$ is still NP-hard: given a set of
points $S$ and a real value $t>1$, it is NP-hard to decide whether a
spanning tree of $S$ with dilation at most $t$ exists.  But for the
special case when the tree is required to be a star, Eppstein and
Wortman~\cite{ew-mds-05} gave an algorithm to compute minimum dilation
stars of a set of points in the plane in $O(n \log n)$ expected time.




In this paper we study the \emph{worst-case behavior} of the function
$\Dil(S,k)$: what is the best dilation one can guarantee for
\emph{any} set $S$ of $n$ points if one is allowed to use $n-1+k$
edges?  In other words, we study the quantity
\[
\dil(n,k) := \sup_{\substack{S \subset \Reals^D \\ |S|=n}} \Dil(S,k).
\]
For the special case when the set $S$ is in $\Reals^2$ and is required
to be in convex position\footnote{A set of points is {\em in convex
position} if they all lie on the boundary of their convex hull.}
we define the quantity $\dil_C(n,k)$ analogously.

The result of Das and Heffernan~\cite{dh-cdsosp-96} mentioned above
implies that, for any constant $\eps'>0$, $\dil(n,\eps'n)$ is
bounded by a constant.
We are interested in what can be achieved for smaller values of $k$,
in particular for $0\leq k< n$.

In the above definitions we have placed the perhaps unnecessary
restriction that the graph may use no other vertices besides the
points of~$S$. We will also consider networks whose vertex sets
are supersets of~$S$. In particular, we define a \emph{Steiner
tree} on $S$ as a tree $T$ with $S \subset V(T)$.
The vertices in $V(T)\setminus S$ are called
\emph{Steiner points}.
Note that $T$ may have any number of vertices, the only restriction
is on its topology.
\paragraph{Our results.}
We first show that any Steiner tree on a set $S$ of $n$ equally
spaced points on a circle has dilation at least $n/\pi$.
We prove in a similar way that $\dil(n,0)\geq \frac{2}{\pi} n-1$.
We remark that Eppstein~\cite{e-sts-00} gave a simpler proof of a
similar bound for $\dil(n,0)$; we improve this bound by a constant
factor.

We then continue with the case $0<k< n$. Here we give an example
of a set $S$ of $n$ points in the plane for which any network with $n-1+k$
edges has dilation at least $\frac{2}{\pi}\lfloor
n/(k+1)\rfloor-1$, proving that $\dil(n,k)\geq
\frac{2}{\pi}\lfloor n/(k+1)\rfloor-1$. We also prove that for
points in convex position the dilation can be almost as large as
in the general case, namely $\delta_C(n,k)=\Omega( n/((k+1)\log
n))$.

Next we study upper bounds. We describe an $O(n\log n)$ time algorithm
that computes for a given set $S$ of $n$ points in $\Reals^D$ and a
parameter $0\leq k< n$ a network of at most $n - 1 + k$ edges, maximum
degree five, and dilation~$O(n/(k+1))$. Combined with our lower
bounds, this implies that $\dil(n,k)= \Theta(n/(k+1))$. In particular,
our bounds apply to the case $k=o(n)$, which was left open by Das and
Heffernan~\cite{dh-cdsosp-96}. Notice that if $k \geq n$, then we have
$1\leq \delta(n,k)\leq \delta(n,n-1)$, and thus
$\delta(n,k)=\Theta(1)$.  This means that $\dil(n,k)= \Theta(1 +
n/(k+1))$ holds for any $k \geq 0$.

Our lower bounds use rather special point sets and it may be the case
that more `regular' point sets admit networks of smaller
dilation. Therefore we also study the special case when $S$ is a point
set with so-called \emph{bounded spread}. The \emph{spread} of a set
of $n$ points $S$ is the ratio between the longest and shortest
pairwise distance in $S$. We show that any set $S$ with spread $s$
admits a network of at most $n - 1 + k$ edges with dilation
$O(s/(k+1)^{1/D})$.  This bound is asymptotically tight for $s =
O(n/(k+1)^{1-1/D})$ (otherwise $O(n/(k+1))$ is a better bound).  This
also leads to tight bounds for the case when $S$ is a $n^{1/D} \times
\cdots \times n^{1/D}$ grid.

\paragraph{Notation and terminology.}
Hereafter $S$ will always denote a set of points in $\Reals^D$.
Whenever it causes no confusion we do not distinguish an edge
$e=(u,v)$ in the network under consideration and the line segment~$uv$.

\section{Lower bounds}
\label{sec:LowerBounds}
In this section we prove lower bounds on the dilation that can be
achieved with $n - 1 + k$ edges, for $0\leq k<n$, by constructing
high-dilation point sets in $\Reals^2$.  Of course, these lower bounds
apply as well in higher dimensions.

\subsection{Steiner trees}
We first show a lower bound on the dilation of any Steiner tree for $S$.
The lower bound for this case uses the set $S$ of $n$ points
$p_1,p_2,\dots,p_n$ spaced equally on
the unit circle, as shown in Fig.~\ref{fig:LBcircle}(a), and is based on
similar arguments as in Ebbers-Baumann et al.~\cite{egk-gdfps-03} (Lemma~3).

\begin{theorem}
  \label{thm:LBsteinerTree}
  For any $n > 1$, there is a set $S$ of $n$ points in convex position such
  that any Steiner tree on $S$ has dilation at
  least~$\frac{1}{\sin(\pi/n)} > \frac{n}{\pi}$.
\end{theorem}
\begin{proof}
  Consider the set $S$ described above
  and illustrated in Fig.~\ref{fig:LBcircle}(a).
  Let $o$ be the center of the circle, and let $T$ be a Steiner tree
  on~$S$. First, we assume that $o$ does not lie on an edge of the
  tree.

  \begin{figure} [htb]
    \centering
    \includegraphics[width=10cm]{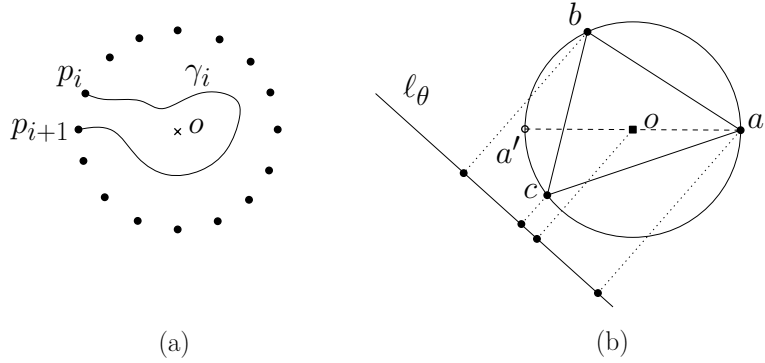}
    \caption{(a) The homotopy class of the path $\gamma_i$.
      (b) Illustrating the proof of Lemma~\ref{lem:triangleLB}.}
    \label{fig:LBcircle}
  \end{figure}

  Let $x$ and $y$ be any two points and let $\gamma$ and $\gamma'$ be
  two paths from $x$ to $y$ avoiding $o$.  We call $\gamma$ and
  $\gamma'$ \emph{(homotopy) equivalent} if $\gamma$ can be deformed
  continuously into $\gamma'$ without moving its endpoints or ever
  passing through the point~$o$, that is, if $\gamma$ and $\gamma'$
  belong to the same homotopy class in the punctured plane
  $\Reals^2\setminus\{o\}$.

  Let $\gamma_i$ be the unique path in $T$ from $p_i$ to $p_{i+1}$
  (where $p_{n+1} := p_1$).  We argue that there must be at least one
  index $i$ for which $\gamma_i$ is not equivalent to the straight segment
  $p_ip_{i+1}$, as illustrated in Fig.~\ref{fig:LBcircle}(a).

  We argue by contradiction. Let $\Gamma$ be the closed loop formed as
  the concatenation of $\gamma_1, \ldots, \gamma_n$, and let $\Gamma'$
  be the closed loop formed as the concatenation of the straight
  segments $p_i p_{i+1}$, for $i = 1\dots n$. If $\gamma_i$ is
  equivalent to $p_i p_{i+1}$, for all $i$, then $\Gamma$ and
  $\Gamma'$ are equivalent.  We now observe that, since $\Gamma'$ is a
  simple closed loop surrounding $o$, it cannot be contracted to a
  point in the punctured plane (formally, it has winding number 1
  around $o$).  On the other hand, $\Gamma$ is contained in the tree
  $T \not\ni o$ (viewed as a formal union\footnote{%
  Notice that $T$ may not be properly embedded in the plane, that is,
  the edges of $T$ may cross or overlap.  However, viewed not as a subset
  of the plane, but rather as an abstract simplicial complex, $T$ is
  certainly simply connected and $\Gamma$ is a closed curve contained
  in it and thus contractible, \emph{within} $T$, to a point.
  Therefore it is also contractible in the punctured plane, as
  claimed.}  of its edges) and hence must be
  contractible in $\Reals^2\setminus\{o\}$.
  Hence $\Gamma$ and $\Gamma'$ cannot be equivalent, a contradiction.

  Consider now a path $\gamma_i$ not equivalent to the segment $p_i
  p_{i+1}$. Then $\gamma_i$ must ``go around''~$o$,
  and so its length is at least~2.
  The distance between $p_i$ and $p_{i+1}$, on the other
  hand, is $2\sin(\pi/n)$, implying the theorem.

  Now consider the case where $o$ lies on an edge of $T$.
  Assume for a contradiction that there is a spanning tree $T$ that has
  dilation $1/\sin(\pi/n) - \eps$ for some $\eps>0$.
  Let $o'$ be a point not on $T$ at distance $\eps/100$ from $o$.
  Then we can use the argument above to show that there are
  two consecutive points
  whose path in $T$ must go around $o'$. By the choice of $o'$ such a path
  must have dilation larger than $1/\sin(\pi/n) - \eps$,
  a contradiction.
\qed
\end{proof}

\subsection{The case $k=0$}

If we require the tree to be a spanning tree without Steiner points,
then the path $\gamma_{i}$ in the above proof must not only ``go
around''~$o$, but must do so using points $p_j$ on the circle only.
We can use this to improve the constant in
Theorem~\ref{thm:LBsteinerTree}, as follows.  Let $p_i$ and $p_{i+1}$
be a pair of consecutive points such that the path $\gamma_i$ is not
equivalent to the segment $p_i p_{i+1}$.  Consider the loop formed by
$\gamma_i$ and $p_{i+1} p_i$.  It consists of straight segments
visiting some of the points of $S$.  Let $C$ be the convex hull of
this loop. The point $o$ does not lie outside $C$ (otherwise, the loop
would be contractible in the punctured plane) and so there exist three
vertices $v_1$, $v_2$, and $v_3$ of $C$ such that $o \in \triangle v_1
v_2 v_3$.  Since the loop visits each of these three vertices once,
its length is at least the perimeter of $\triangle v_1 v_2 v_3$, which
is at least 4 by Lemma~\ref{lem:triangleLB} below. Therefore, we have
proven


\begin{corollary}
  \label{cor:LBtree}
  For any $n>1$, $$\dil_C(n,0) \geq \frac{4 -
  2\sin(\pi/n)}{2\sin(\pi/n)} \geq\frac{2n}{\pi} -1.$$
\end{corollary}

\begin{lemma} \label{lem:triangleLB}
  Any triangle inscribed in a unit circle and containing the circle center
  has perimeter at least~$4$.
\end{lemma}
\begin{proof}
  Let $o$ be the circle center, and let $a$, $b$, and $c$ be three points at
  distance one from $o$ such that $o$ is contained
  in the triangle~$\triangle abc$.
  We will prove that the perimeter $p(\triangle abc)$ is at least~$4$.

  We need the following definition: For a compact convex set $C$ in
  the plane and $0 \leq \theta < \pi$, let $w(C, \theta)$ denote the
  \emph{width of $C$ in direction $\theta$}.  More precisely, if
  $\ell_\theta$ is the line through the origin with normal vector
  $(\cos \theta, \sin \theta)$, then $w(C,\theta)$ is the length of
  the orthogonal projection of $C$ to $\ell_\theta$.  The
  Cauchy-Crofton formula~\cite{c-dgcs-76} allows us to express the
  perimeter $p(C)$ of a compact convex set $C$ in the plane as $p(C) =
  \int_{0}^{\pi} w(C,\theta) d\theta$.

  We apply this formula to $\triangle abc$, and consider its
  projection on the line $\ell_\theta$ (for a given~$\theta$).  We
  choose a coordinate system where $\ell_\theta$ is horizontal.  By
  mirroring and renaming the points $a, b, c$, we can assume that $b$
  is the leftmost point, $a$ is the rightmost point, that $o$ lies
  below the edge~$ab$, and that $c$ does not lie to the right of~$o$.
  This immediately implies that $w(oc,\theta) \leq w(ob,\theta)$.
  Consider now the point $a'$ obtained by mirroring $a$ at~$o$.  Since
  $c$ must lie below the line $aa'$ and not left of $b$, we have
  $w(oc,\theta) \leq w(oa',\theta) = w(oa,\theta)$.  This implies
  $3w(oc,\theta) \leq w(oa,\theta) + w(ob,\theta) + w(oc,\theta)$, and
  therefore
  \[
  \begin{array}{lll}
    w(\triangle abc,\theta) & = & w(oa,\theta) + w(ob,\theta) \\
    & = & w(oa,\theta) + w(ob,\theta) + w(oc,\theta) - w(oc,\theta) \\
    & \geq & \frac 2 3(w(oa,\theta) + w(ob,\theta) + w(oc,\theta)).
  \end{array}
  \]
  Integrating $\theta$ from $0$ to $\pi$ and applying the
  Cauchy-Crofton
  formula gives $p(\triangle abc) \geq \frac 2 3 (p(oa) + p(ob)
  + p(oc))$.  Since $oa$, $ob$, and $oc$ are segments of length~1, each
  has perimeter 2, and so we have $p(\triangle abc) \geq \frac 2 3 \cdot
  6 = 4$.
  \qed
\end{proof}

\subsection{The general case}
We now turn to the general case, and we consider graphs with $n - 1 + k$
edges for $0<k<n$.


\begin{theorem} \label{thm:LBk}
  For any $n$ and any $k$ with $0< k < n$,
$$\dil(n,k) \geq \frac{2}{\pi} \cdot \Big \lfloor \frac{n}{k+1}\Big
\rfloor-1.$$
\end{theorem}

\begin{proof}
  Our example $S$ consists of $k + 1$ copies of the set used in
  Theorem~\ref{thm:LBsteinerTree}.  More precisely, we choose sets $S_i$, for
  $1\leq i\leq k+1$, each consisting of
  at least $\lfloor n/(k+1)\rfloor$ points.
  We place the points in $S_i$ equally spaced on
  a unit-radius circle with center at $(2 n i, 0)$, as in
  Fig.~\ref{fig:LBk}. The set $S$ is the union of $S_1, \ldots ,
  S_{k+1}$; we choose the cardinalities of the $S_i$ such that $S$
  contains $n$~points.
  \begin{figure} [htb]
    \centering
    \includegraphics[width=12cm]{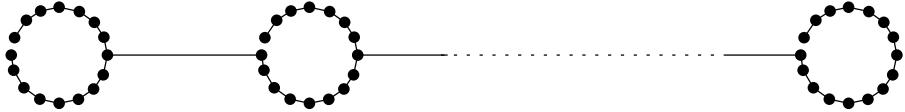}
    \caption{Illustrating the point set $S$ constructed in the proof
      of Theorem~\ref{thm:LBk}.}
    \label{fig:LBk}
  \end{figure}

  Let $G$ be a graph with vertex set~$S$ and $n - 1 + k$ edges.  We
  call an edge of $G$ \emph{short} if its endpoints lie in the same
  set $S_i$, and \emph{long} otherwise. Since $G$ is connected, there
  are at least $k$ long edges, and therefore at most $n-1$ short
  edges. Since $\sum |S_i|=n$, this implies that there is a set $S_i$
  such that the number
  of short edges with both endpoints in $S_i$ is at most $|S_i|-1$.
  Let $G'$ be the induced
  subgraph of~$S_{i}$.  If $G'$ is not connected, then $S_{i}$
  contains two points connected in $G$ using a long edge, and
  therefore with dilation at least~$n$.  If $G'$ is connected, then it
  is a tree, and so the argument of
  Corollary~\ref{cor:LBtree} implies that its dilation is
  at least
  \[
  \frac{2 - \sin(\pi / \lfloor n/(k+1)\rfloor) }
  {\sin(\pi/\lfloor n/(k+1)\rfloor)}
  \geq \frac{2}{\pi} \cdot \left\lfloor \frac{n}{k+1}\right\rfloor-1 .
  \]
  This implies the claimed lower bound on the dilation of~$G$.
  \qed
\end{proof}

\subsection{Points in convex position}
\begin{figure} [htb]
  \centering
  \includegraphics[width=12cm]{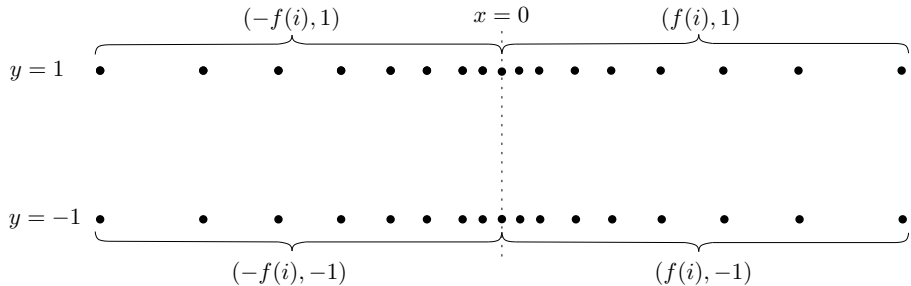}
  \caption{Illustrating the lower bound construction used in
    the proof of Theorem~\ref{thm:LB_convex} with $n=34$.}
  \label{fig:convexLB}
\end{figure}

The point set of Theorem~\ref{thm:LBsteinerTree} is in convex
position, but works as a lower bound only for $k = 0$. In fact, by
adding a single edge (the case $k=1$) one can reduce the dilation
to a constant. Now consider $n$ points that lie on the boundary of
a planar convex figure with aspect ratio at most $\rho$, that is,
with the ratio of diameter to width at most~$\rho$. It is not
difficult to see that connecting the points along the
boundary---hence, using $n$ edges---leads to a graph with dilation
$\Theta(\rho)$. However, the following theorem shows that for
large aspect ratio, one cannot do much better than in the general
case.
\begin{theorem}
  \label{thm:LB_convex}
  For any $n$ and any $k$ with $0 \leq k < n$,
   $$\dilConv(n,k) = \Omega \Big (\frac n {(k+1)\log n}\Big).$$
\end{theorem}
\begin{proof}
  We present the proof for $k=0$, but the construction can be
  generalized to hold for any $k>0$ by using the same idea as in
  the proof of Theorem~\ref{thm:LBk}: placing $k+1$ copies of the
  construction along a horizontal line, with sufficient space
  between consecutive copies.

  For the case when $k=0$, set $m = \lfloor n/4 - 1/2\rfloor$ and let
  $o := (0,0)$.
  Consider the function $f(i) = (1 + \frac{\ln m}{m})^{i} - 1$.
  Our construction
  consists of the $4m + 2 \leq n$ points $S$ with coordinates $(f(i), 1)$,
  $(f(i), -1)$, $(-f(i),1)$, and $(-f(i), -1)$, for $i \in \{0,\ldots,m\}$,
  as shown in Fig.~\ref{fig:convexLB}.
  The points of $S$ lie on the boundary of a rectangle, so $S$ is
  in convex position. (A slight perturbation of $S$ would even
  give a set in strictly convex position, that is, a set where every
  point is extreme.)
  Consider a minimum-dilation tree $T$ of $S$. As in
  the proof of Lemma~\ref{thm:LBsteinerTree}, we can now argue that
  there must exist two \emph{consecutive} points $p$ and $q$ in $S$ such that
  the path in $T$ connecting them is not (homotopy) equivalent
  to the straight-line segment between them in the punctured
  plane $\Reals^2{\setminus}\{o\}$.
  By symmetry, we can assume that $p$ and $q$ lie in the half-plane $x
  \geq 0$, and that $p$ lies above the $x$-axis.

  The first case is $p=(f(i),1)$ and $q=(f(i+1),1)$, where $0 \leq i <
  m$. The Euclidean distance between them is $f(i+1) - f(i) =
  \frac{\ln n}{n}(1+f(i))$, and the length of the shortest path in $T$
  is at least $\sqrt{2}(1 + f(i))$.
  The two bounds imply that the dilation of $T$ is at least $\sqrt{2} \frac
  {n} {\ln n}$ in this case.

  The second case is $p=(f(m),1)$ and $q=(f(m),-1)$.  Then the
  Euclidean distance between them is~$2$, and the length of the
  shortest path in $T$ between them is at least $2 f(m)$, thus $T$ has
  dilation $f(m)$ in this case. It remains to bound $f(m)$, which can
  be done by using the inequality $(1+t/m)^{m} \geq e^{t}(1 -
  t^{2}/m)$, which holds for $|t| \leq
  m$~\cite[Proposition~B.3]{mr-ra-95}.  We obtain
  \[
  f(m) = \Big(1+\frac{\ln m}{m}\Big)^{m} - 1 \geq m
  \Big(1-\frac{\ln^{2}m}{m}\Big) - 1 = \Omega(m),
  \]
  which concludes the proof of the theorem.  \qed
\end{proof}

\section{A constructive upper bound}
\label{sec:ApxAlgorithm}

In this section we show an upper bound on the dilation achievable with
$k$ extra edges.  We make use of the fact (observed by
Eppstein~\cite{e-sts-00}) that a minimum spanning tree has dilation at
most~$n-1$. We will use the following lemma.
\begin{lemma}
  \label{lem:mst}
  Let $S$ be a set of $n$ points in $\Reals^D$, and let $T$ be a
  minimum spanning tree of $S$. Then $T$ has dilation at most $n-1$.
  Hence $\Dil(S,0) \leq \Dil(T) \leq n-1$
  and, as this holds for all $S$ with $|S|=n$,
  we have $\dil(n,0) \leq n-1$.
\end{lemma}
\begin{proof}
  Let $p, q \in S$ and consider the path $\gamma$ connecting $p$ and
  $q$ in~$T$.  Since $T$ is a minimum spanning tree, any edge
  in~$\gamma$ has length at most~$d(p,q)$.  Since $\gamma$ consists
  of at most $n-1$ edges, the dilation of $\gamma$ is at most~$n - 1$.
  \qed
\end{proof}

\subsection{The planar case, $D=2$}
\label{sec:planarupper}

The following algorithm builds a spanner with at most $n - 1 + k$ edges:
\begin{algorithm}
\caption{\textsc{SparseSpanner}($S, k$)}
\label{alg:Algorithm}
\newcommand{\RETURN}{\STATE\textbf{return} }
\begin{algorithmic}[1]
\REQUIRE{A set $S$ of $n$ points in the plane and an integer
  $k \in \{0,\ldots,2n-5\}$.}
\ENSURE{A graph $G=(S,E)$ with dilation $O(n/(k+1))$ and at most $n-1+k$ edges.}
\STATE Compute a Delaunay triangulation of $S$.
\STATE Compute a minimum spanning tree $T$ of $S$.
\IF {$k=0$}
\RETURN $T$.
\ENDIF
\STATE Let $m \leftarrow \lfloor (k+5)/2 \rfloor$.
\STATE Compute $m$ disjoint subtrees of $T$,
each containing $\O(n/m)$ points, by removing $m-1$ edges.
\STATE $E \leftarrow \emptyset$.
\FOR {each subtree $T'$}
\STATE add the edges of $T'$ to $E$.
\ENDFOR
\FOR {each pair of subtrees $T'$ and $T''$}
\IF {there is a Delaunay edge $(p,q)$ with $p \in T'$, $q \in T''$}
\STATE add the shortest such edge $(p,q)$ to $E$.\ENDIF\ENDFOR
\RETURN $G=(S,E)$.
\end{algorithmic}
\end{algorithm}

\begin{figure} [htb]
  \centering
  \includegraphics[width=12cm]{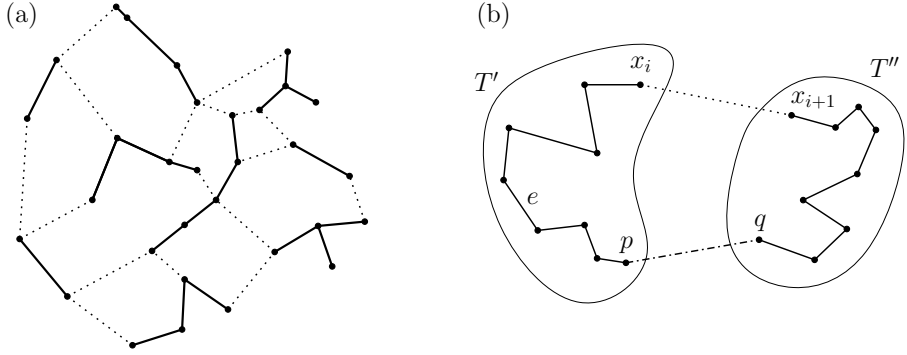}
  \caption{(a) The solid edges connect vertices within a subtree,
    while the dotted edges connect different subtrees. 
    (b) Illustrating the proof of Lemma~\ref{lem:algCorrect}.}
  \label{fig:lemma3}
\end{figure}

We first prove the correctness of the algorithm.
\begin{lemma}
  \label{lem:algCorrect}
  Algorithm \textsc{SparseSpanner} returns a graph $G$ with at most $n
  - 1 + k$ edges and dilation  bounded by $O(n/(k+1))$.
\end{lemma}
\begin{proof}
  Lemma~\ref{lem:mst} shows that our algorithm is correct if $k=0$, so
  from now on we assume that $k \geq 1$, and thus $m \geq 3$.  The
  output graph~$G$ is a subset of a Delaunay triangulation of~$S$ and
  is therefore a planar graph, see Fig.~\ref{fig:lemma3}a.  Consider 
  now the graph $G'$ obtained
  from~$G$ by contracting each subtree $T'$ created in step~6 to a
  single node.  Since $G$ is planar and since two subtrees are connected 
  with at most one edge, $G'$~is a planar graph with $m \geq 3$ vertices, 
  without loops or multiple edges, and so it has at most $3m - 6$ edges. 
  The total number of edges in the output graph is therefore at most
  \[(n-1-(m-1))+(3m-6)=n+2m-6 \leq n+k-1.
  \]

  Next we prove the dilation bound. Consider two points $x, y \in S$.
  Let $x = x_{0},x_{1},\dots,x_{j} = y$ be a shortest path from $x$ to
  $y$ in the Delaunay graph. The dilation of this path is bounded by
  ${2\pi}/({3 \cos(\pi/6)}) = O(1)$~\cite{kg-cgwac-92}.  We claim that
  each edge $x_{i}x_{i+1}$ can be replaced by a path in $G$ with
  dilation $O(n/k)$.  The concatenation of these paths yields a path
  from $x$ to $y$ with dilation~$O(n/k)$, proving the lemma.

  If $x_{i}$ and $x_{i+1}$ fall into the same subtree $T'$, then
  Lemma~\ref{lem:mst} implies a dilation of $O(n/m) = O(n/k)$.

  It remains to consider the case $x_{i} \in T'$, $x_{i+1} \in T''$,
  where $T'$ and $T''$ are distinct subtrees of $T$, see 
  Fig.~\ref{fig:lemma3}b.  Let $(p, q)$ be
  the edge with $p\in T'$, $q\in T''$ inserted in step~12---such an
  edge exists, since there is at least one Delaunay edge between $T'$
  and $T''$, namely $(x_{i},x_{i+1})$.  Let $\mu'$ be a path from
  $x_{i}$ to $p$ in $T'$, and let $\mu''$ be a path from $q$ to
  $x_{i+1}$ in $T''$.  Both paths have $O(n /k)$ edges.
  By construction, we have $d(p,q) \leq d(x_{i},x_{i+1})$.  We claim
  that every edge $e$ on $\mu'$ and $\mu''$ has length at
  most~$d(x_{i},x_{i+1})$.  Indeed, if $e$ has length larger than
  $d(x_{i},x_{i+1})$, then the original tree $T$ (of which $T'$ and
  $T''$ are subtrees) cannot be a minimum spanning tree of~$S$: we can
  remove $e$ from $T$ and insert either $(x_{i},x_{i+1})$ or $(p,q)$
  to obtain a better spanning tree.
  Therefore the concatenation of $\mu'$, the edge $(p,q)$,
  and $\mu''$ is a path of $O(n/k)$ edges, each of length at most
  $d(x_{i},x_{i+1})$, and so the dilation of this path is~$O(n/k)$.
  \qed
\end{proof}


\begin{theorem}\label{th:2Dspanner}
  Given a set $S$ of $n$ points in the plane and an integer $0 \leq k
  \leq 2n-5$, a graph $G$ with vertex set $S$, $n-1+k$ edges, and
  dilation $O(n/(k+1))$ can be constructed in $O(n \log n)$ time.
\end{theorem}
\begin{proof}
  We use algorithm \textsc{SparseSpanner}. Its correctness has been
  proven in Lemma~\ref{lem:algCorrect}.
  Steps 1~and~2 can be implemented in $O(n \log n)$
  time~\cite{e-sts-00}.
  Step~6 can be implemented in linear time as follows.  Orient $T$ by
  choosing any node of degree at most five as root node.  Since the
  degree of a minimum spanning tree is at most
  six~\cite{e-sts-00,y-cmstk-82}, every node in $T$ has at most five
  children.  We will partition $T$ into disjoint subtrees by removing
  edges, such that each subtree has size at least $n/m$ and at most
  $5(n/m) - 4$.  The only exception is the top subtree---that is, the
  subtree containing the root $r$ of $T$---which has size at most
  $5(n/m) - 4$ but could be smaller than $n/m$.  To obtain this
  partition, we first partition the trees rooted at the at most five
  children of $r$ recursively.  All subtrees we get in this manner will
  have sizes between $n/m$ and $5(n/m) - 4$, except possibly for the
  subtrees containing the children of $r$ which could be too
  small. The subtrees that are too small are grouped together with $r$
  into one new subtree of size at most $5(n/m) - 4$. Since all but one
  subtree have size at least $n/m$, the number of subtrees is at most
  $m$. To obtain exactly $m$ subtrees, we can simply remove some more
  edges arbitrarily.
  To implement steps~10--12, we scan the edges of the Delaunay
  triangulation, and keep the shortest edge connecting each pair of
  subtrees. This can be done in $O(n \log n)$ time.
  \qed
\end{proof}

\subsection{The higher-dimensional case, $D \geq 2$}
\label{sec:higher}
The algorithm of Section~\ref{sec:planarupper} uses the fact that the
Delaunay triangulation has constant dilation, a linear number of
edges, and can be computed efficiently.  These properties do not
generalize to dimensions $D \geq 3$, and so we have to turn to another
bounded-degree spanner of our point set~$S$.  Since the minimum
spanning tree is unlikely to be computable in near-linear time in
dimension $D \geq 3$, we will use a minimum spanning tree \emph{of
this spanner} instead of a minimum spanning tree of the point set
itself.  These two ideas will allow us to generalize our result to any
constant dimension $D \geq 2$. Moreover, we show that this can
be achieved by a graph with degree at most five.

\paragraph{Properties of the minimum spanning tree of a spanner.}
Let $t \geq 1$ be a real number, let $G$ be an arbitrary $t$-spanner
for $S$, and let $T$ be a minimum spanning tree of $G$. In the
following three lemmas, we prove that $T$ has ``approximately'' the
same properties as an exact minimum spanning tree of the point set~$S$.

\begin{lemma}  \label{lemma1}
  Let $p$ and $q$ be two points of $S$. Then every edge on the
  path in $T$ between $p$ and $q$ has length at most
  $t \cdot d(p,q)$.
\end{lemma}
\begin{proof}
Let $(x,y)$ be an arbitrary edge on the path in $T$ from $p$
to~$q$.  For a contradiction, assume that $d(x,y) > t \cdot d(p,q)$.

Let $T'$ be the graph obtained by removing the edge $(x,y)$ from~$T$.
It consists of two components, one containing $p$, the other
containing $q$.  Now, since $G$ is a $t$-spanner, it contains a path
$\gamma$ between $p$ and $q$ of length at most $t \cdot d(p,q)$.  The union
of $T'$ with the path $\gamma$ is a spanning graph of $G$, and its weight
is less than the weight of~$T$, a contradiction.  \qed
\end{proof}

\begin{lemma}   \label{lemma2}
  $T$ is a $t(n-1)$-spanner for $S$.
\end{lemma}
\begin{proof}
Let $p$ and $q$ be distinct points of $S$, and let $\gamma$ be the
path in $T$ between $p$ and $q$. By Lemma~\ref{lemma1}, each edge
of $\gamma$ has length at most $t \cdot d(p,q)$. Since $\gamma$ contains at most
$n-1$ edges, it follows that the length of $\gamma$ is at most
$t(n-1) \cdot d(p,q)$.
\qed
\end{proof}

\begin{lemma}   \label{lemma4}
       Let $m$ be an integer with $1 \leq m \leq n-1$, and let $T'$
       and $T''$ be two vertex-disjoint subtrees of $T$, each
       consisting of at most $m$ vertices. Let $p$ be a vertex of
       $T'$, let $q$ be a vertex of $T''$, and let $\gamma$ be the path
       in $T$ between $p$ and $q$. If $x$ is a vertex of $T'$ that is
       on the subpath of $\gamma$ within $T'$, and $y$ is a vertex of $T''$
       that is on the subpath of $\gamma$ within $T''$, then
       \[ d(x,y) \leq ( 2t(m-1) + 1 ) \cdot d(p,q) .
       \]
\end{lemma}
\begin{proof}
Let $\gamma'$ be the subpath of $\gamma$ between $p$ and $x$. By
Lemma~\ref{lemma1}, each edge of $\gamma'\subset \gamma$ has length at most
$t \cdot d(p,q)$.
Since $\gamma'$ contains at most $m-1$ edges, it follows that this path
has length at most $t(m-1) \cdot d(p,q)$. On the other hand, since
$\gamma'$ is a path between $p$ and $x$, its length is at least
$d(p,x)$. Thus, we have $d(p,x) \leq t(m-1) \cdot d(p,q)$. A symmetric
argument gives
$d(q,y) \leq t(m-1) \cdot d(p,q)$. Therefore, we have
\begin{eqnarray*}
d(x,y) & \leq & d(x,p) + d(p,q) + d(q,y) \\
       & \leq & t(m-1) \cdot d(p,q) + d(p,q) + t(m-1) \cdot d(p,q) ,
\end{eqnarray*}
completing the proof of the lemma.
\qed
\end{proof}

\paragraph{A graph with at most $n - 1 + k$ edges and dilation $O(n/(k+1))$.}
Fix a constant $t>1$.
Let $c$ be a constant such that for any set $X$ of $2m$ points, the
$t$-spanner of Das and Heffernan~\cite{dh-cdsosp-96} has at most $cm$ edges.
Let $G$ be a $t$-spanner for $S$ whose degree is bounded by a constant.
Since then the minimum spanning tree $T$ of $G$ has bounded
degree as well, we can use the algorithm described in the proof of
Theorem~\ref{th:2Dspanner} (step~6) to remove $m = \lfloor
k/(c-1)\rfloor$ edges from $T$ and obtain vertex-disjoint subtrees
$T_0,T_1,\ldots,T_m$, each containing $O(n/(k+1))$ vertices.
The vertex sets of these subtrees form a partition of $S$. Let $X$ be the
set of endpoints of the $m$ edges that are removed from $T$.  The
size of $X$ is at most $2m$.

We define $G'$ to be the graph with vertex set $S$ that is the union of
\begin{enumerate}
\item the trees $T_0,T_1,\ldots,T_m$, and
\item Das and Heffernan's $t$-spanner $G''$ for the set $X$ ($G''$ is
  empty if $m = 0$).
\end{enumerate}
We first observe that the number of edges of $G'$ is bounded from
above by $n - 1 - m + cm \leq n - 1 + k$.

\begin{lemma} \label{lemma5}
       The graph $G'$ has dilation $O(n/(k+1))$.
\end{lemma}
\begin{proof}
The statement follows from Lemma~\ref{lemma2} if $m = 0$. Let $m > 0$,
and let $p$ and $q$ be two distinct points of $S$. Let $i$ and $j$ be
the indices such that $p$ is a vertex of the subtree $T_i$ and $q$ is
a vertex of the subtree $T_j$.

First assume that $i=j$. Let $\gamma$ be the path in $T_i$ between $p$
and $q$. Then, $\gamma$ is a path in $G'$. By Lemma~\ref{lemma1},
each edge on $\gamma$ has length at most $t \cdot d(p,q)$. Since $T_i$
contains $O(n/k)$ vertices, the number of edges on $\gamma$ is $O(n/k)$.
Therefore, since $t$ is a constant, the length of $\gamma$ is
$O(n/k) \cdot d(p,q)$.

Now assume that $i \neq j$. Let $\gamma$ be the path in $T$ between $p$
and $q$. Let $(x,x')$ be the edge of $\gamma$ for which $x$ is a
vertex of $T_i$, but $x'$ is not a vertex of $T_i$. Similarly,
let $(y,y')$ be the edge of $\gamma$ for which $y$ is a vertex of $T_j$,
but $y'$ is not a vertex of $T_j$. Then, both $(x,x')$ and
$(y,y')$ are edges of $T$ that have been removed when the
subtrees were constructed. Hence, $x$ and $y$ are both contained
in $X$ and, therefore, are vertices of $G''$.
Let $\gamma_i$ be the path in $T_i$ between $p$ and $x$, let $\gamma_{xy}$ be
a shortest path in $G''$ between $x$ and $y$, and let $\gamma_j$ be the
path in $T_j$ between $y$ and $q$. The concatenation $\gamma'$ of $\gamma_i$,
$\gamma_{xy}$, and $\gamma_j$ is a path in $G'$ between $p$ and $q$.

Since both $\gamma_i$ and $\gamma_j$ are subpaths of $\gamma$, it follows from
Lemma~\ref{lemma1} that each edge on $\gamma_i$ and $\gamma_j$ has length at
most $t \cdot d(p,q)$. Since $T_i$ and $T_j$ contain $O(n/k)$ vertices,
it follows that the sum of the lengths of $\gamma_i$ and $\gamma_j$ is
$O(n/k) \cdot d(p,q)$. The length of $\gamma_{xy}$ is at most
$t \cdot d(x,y)$ which, by Lemma~\ref{lemma4}, is also
$O(n/k) \cdot d(p,q)$. Thus, the length of $\gamma'$ is
$O(n/k) \cdot d(p,q)$.
\qed
\end{proof}

Now let $G$ be the $t$-spanner of Das and Heffernan~\cite{dh-cdsosp-96}.
This spanner can be computed in $O(n \log n)$ time, and each vertex
has degree at most three. Given $G$, its minimum spanning tree $T$
and the subtrees $T_0,T_1,\ldots,T_m$
can be computed in $O(n \log n)$ time. Finally, $G''$ can be computed
in $O(m \log m) = O(n \log n)$ time, and each vertex has degree at
most three. Thus $G'$ has dilation $O(n/(k+1))$, it contains at most
$n-1+k$ edges, and it can be computed in $O(n \log n)$ time.

We analyze the degree of $G'$: Consider any
vertex $p$ of $G'$. If $p \not\in X$, then the degree of $p$ in
$G'$ is equal to the degree of $p$ in $T$, which is at most three.
Assume that $p \in X$. The graph $G''$ contains at most three edges
that are incident to $p$. Similarly, the tree $T$ contains at most
three edges that are incident to $p$, but, since $p \in X$, at least
one of these three edges is not an edge of $G'$. Therefore, the
degree of $p$ in $G'$ is at most five. Thus, each vertex of $G'$ has
degree at most five.

Thus, we have proved the following result:

\begin{theorem}   \label{thmmain}
  Given a set $S$ of $n$ points in $\Reals^D$ and an integer $k$ with
  $0 \leq k < n$, a graph $G$ with vertex set $S$, $n-1+k$ edges,
  degree at most five, and dilation $O(n/(k+1))$ can be constructed in
  $O(n \log n)$ time.
\end{theorem}

\section{Bounded Spread}

In this section we consider the case when the set of input points has
bounded spread. The \emph{spread} of a set of $n$ points $S$, denoted
$s(S)$, is the ratio between the longest and shortest pairwise
distance in $S$. In $\Reals^{D}$ we have $s(S)=\Omega(n^{1/D})$.

We define the function
\[
\dil(n,s,k) := \sup \{ \Dil(S,k) \ | \ S\subset \Reals^D, \,|S|=n, \,
s(S)\leq s \},
\]
that is, the best dilation one can guarantee for any
set $S$ of $n$ points with spread $s$ if one is allowed to use
$n-1+k$ edges.
\begin{theorem} \label{thm:spread_UB}
  For any $n$, any $s$, and any $k$ with $0 \leq k < n$,
  \[
  \delta(n,s,k)= O\bigl(s/(k+1)^{1/D}\bigr).
  \]
\end{theorem}
\begin{proof}
  Assume without loss of generality that the smallest interpoint
  distance in $S$ is 1.

  The case $k \leq 2$ is nearly trivial: we pick an arbitrary point $u
  \in S$ and connect all other points to $u$.  Since the shortest
  interpoint distance is~$1$, the dilation is at most~$2s$.

  For the case $k > 2$, let $B$ be an axis-parallel cube with side
  length~$s$ containing~$S$, and let $m := \lfloor
  (k-1)^{1/D}\rfloor$.  We partition $B$ into $m^D$ small cubes with
  side length~$s/m$. Let us call each small cube a \emph{cell}.  Let
  $X \subset S$ be a set of \emph{representative points}, that is, one
  point of $S$ taken from each non-empty cell.  Let $G'$ be a
  $t$-spanner for $X$ with $2|X|$ edges and constant~$t$~\cite{dh-cdsosp-96}.
  We obtain our final graph $G$ from $G'$ by connecting each point $p
  \in S$ to the representative point of the cell containing~$p$.

  Since $|X| \leq m^{D} \leq k - 1$, the total number of edges of $G$
  is at most $n - |X| + 2|X| \leq n - 1 + k$.

  It remains to prove that the dilation of $G$ is $O(s/m)$ for every
  two points $p,q\in S$. If $p$ and $q$ lie in the same cell, then
  the bound follows immediately.
  If $p$ and $q$ lie in different cells, then let $p'$ and $q'$ be
  their representative points.  The length of the shortest path
  between $p$ and $q$ in $G$ is then
  \begin{align*}
    d_G(p,q) & = d_G(p,p')+d_G(p',q')+d_G(q',q) \\
    & \leq d(p,p') + t \cdot d(p', q') + d(q', q)\\
    & \leq d(p,p') + t \cdot (d(p', p) + d(p,q) + d(q, q')) + d(q', q)\\
    & = (1 + t)(d(p, p') + d(q', q)) + t \cdot d(p,q)\\
    & \leq (1 + t)2\sqrt{D} \frac {s}{m}+ t \cdot d(p,q)\\
    & \leq \bigl((1 + t)2\sqrt{D} \frac {s}{m}+ t \bigr) \cdot d(p,q).
  \end{align*}
  The last inequality uses $d(p,q) \geq 1$.
  \qed
\end{proof}
To prove a corresponding lower bound, we need a lemma in the spirit of
Theorem~\ref{thm:LBsteinerTree}.
\begin{lemma}
  \label{lem:LBsquare}
  Let $Q$ be a two-dimensional square of side length~$2\sigma$ in
  $\Reals^{D}$, let $S$ be a set of points on the boundary of $Q$, and
  let $T$ be a Steiner tree for $S$ (with vertices in $\Reals^{D}$).
  Then there are two consecutive points $p,\,q\in S$ such that
  $d_{T}(p,q) \geq 2\sigma$.
\end{lemma}
\begin{proof}
  Let $T'$ be the orthogonal projection of $T$ onto the plane spanned
  by~$Q$.  Edges of $T$ may cross or overlap in the projection,
  vertices may even coincide, but viewed as a formal union of its
  edges, $T'$ is now a Steiner tree for $S$.  The argument made in the
  proof of Theorem~\ref{thm:LBsteinerTree} therefore goes through;
  there are consecutive points $p$, $q$ in $S$ such that the path from
  $p$ to $q$ in $T'$ goes around the center of $Q$, and therefore its
  length is at least~$2\sigma$.
  \qed
\end{proof}

\begin{figure} [htb]
  \centering
  \includegraphics[width=12cm]{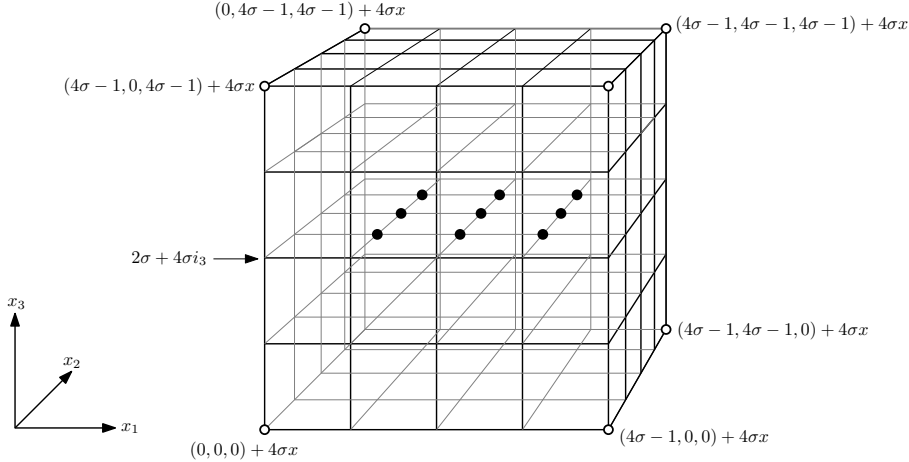}
  \caption{The cube illustrates the cell of an integer vector $x$ in 3
  dimensions and the set of solid black points illustrates the square
  of the cell.}
  \label{fig:theorem7}
\end{figure}

\begin{theorem}
  \label{thm:spread_LB}
  For any integer $r \geq 4$, any integer $m$ with $1 \leq m \leq r/4$, and
  any integer $n$ with $2r m^{D-1} \leq n \leq r^{D}$, there
  is a subset $S$ of $n$ points of the $r\times r \times\cdots \times
  r$ integer grid $\{0, 1, \dots, r-1\}^{D}$ such that
  \[
  \delta(S,m^{D} - 1) \geq \frac{r}{2m} - 1.
  \]
\end{theorem}
\begin{proof}
  Let $\sigma := \lfloor r/(4m)\rfloor$.  We define $m^{D}$ small
  subgrids as follows: For an integer vector $x = (i_{1}, i_{2},
  \dots, i_{D}) \in \{0,\dots,m-1\}^{D}$, let the \emph{cell} $C(x) :=
  \{0, 1, \dots, 4\sigma - 1\}^{D} + 4\sigma x$.  For cell $C(x)$, we
  define its \emph{square} $Q(x) := (\{\sigma, \dots, 3\sigma\}^{2}
  \times \{2\sigma\}^{D-2}) + 4\sigma x$, as illustrated in 
  Fig.~\ref{fig:theorem7}.  Note that cells are
  $D$-dimensional and contain $4^{D}\sigma^{D}$ points each, squares are
  two-dimensional and contain $(2\sigma+1)^{2}$ points each.  We are in fact
  only interested in the boundary points of each square, there are
  $8\sigma$ such boundary points in each square, and therefore
  $8m^{D}\sigma$ such points in total.  Our set $S$ contains all these
  square boundary points, and then an arbitrary subset of the
  remaining grid points~$\{0, \dots, r-1\}^{D}$.

  It remains to prove that any network $G$ with $n-1+(m^{D} - 1)$
  edges on $S$ has dilation at least~$2\sigma$.  Let us call an edge
  $e$ of $G$ \emph{short} if both endpoints of $e$ are in the same
  cell, and \emph{long} otherwise.  Let $T$ be any spanning tree
  of~$G$. We color the $m^{D} - 1$ edges of $G$ that are not in~$T$
  red.  Since there are at most $m^{D} - 1$ short red edges in $G$ and
  $m^{D}$ cells, there must be a cell $C(x)$ without a short red edge.
  We apply Lemma~\ref{lem:LBsquare} to the boundary points $S'$ of the
  square $Q(x)$ inside $C(x)$. The tree $T$ is a Steiner tree for
  $S'$, and so there are consecutive points $p,\, q \in S'$ such that
  $d_{T}(p,q) \geq 2\sigma$.  If $d_{G}(p, q) = d_{T}(p,q)$, then this
  implies $\Delta(G) \geq 2\sigma$.  Otherwise, the shortest path from
  $p$ to $q$ in $G$ must use a red edge.  Since $C(x)$ contains no
  short red edge, the path must therefore pass through a point outside
  $C(x)$, which implies that it has length at least $2\sigma$, and so
  $\Delta(G) \geq 2\sigma$.  \qed
\end{proof}

Combining Theorems~\ref{thm:spread_UB} and~\ref{thm:spread_LB}, we
obtain an asymptotically sharp bound for the dilation.
\begin{corollary}
  For any $n$, any $k$ with $0 \leq k < n$, and any $s$ with $s =
  \Omega(n^{1/D})$ and $s = O(n/(k+1)^{1-1/D})$, we have
  \[
  \delta(n,s,k)= \Theta\bigl(s/(k+1)^{1/D}\bigr).
  \]
\end{corollary}
\begin{proof}
  The upper bound is Theorem~\ref{thm:spread_UB}.  For the lower
  bound, set $r := \lceil s/\sqrt{D} \rceil$ and $m := \lceil
  (k+1)^{1/D} \rceil$, apply Theorem~\ref{thm:spread_LB} and observe
  that the resulting grid set has spread at most~$s$.
  \qed
\end{proof}

The ``most regular'' point set one might imagine is the regular
grid. It turns out that the spread-based lower bound remains true even
for this set.

\begin{corollary}
  \label{thm:grid}
  Let $\G$ be a set of $n = r^{D}$ points forming a $r \times r \times
  \cdots \times r$-grid, and let $0 \leq k < n$.  Then
  \[
  \Delta(\G, k) = \Theta\bigl(r/(k+1)^{1/D}\bigr).
  \]
\end{corollary}
\begin{proof}
  The upper bound follows from the fact that $\G$ has spread
  $s=\Theta(r)$ (better constants can be obtained by a direct
  construction).  The lower bound follows from
  Theorem~\ref{thm:spread_LB}, by observing that if you set $n$ to
  $r^{D}$, then the construction results in $S = \G$.  \qed
\end{proof}


\section{Open problems}
\label{sec:conclusions}
We have shown that for any $n$-point set $S$ in $\Reals^D$ and any
parameter $0 \leq k < n$, there is a graph $G$ with vertex set $S$, $n
- 1 + k$ edges, degree at most five, and dilation $O(n/(k+1))$. We
also proved a lower bound of $\Omega(n/(k+1))$ on the maximum dilation
of such a graph.  An interesting open problem is whether the degree
can be reduced to four, or even three.

The constant in our lower bound for $\dil(n/k)$ is $2/\pi$, but we
have only proven asymptotically matching upper bounds.  Even for the
case $k=0$, the upper bound is $n-1$ while the lower bound is only
$2n/\pi - 1$, and it would be interesting to establish the right
constant.  For $k > 0$ the discrepancy between upper and lower bounds
is even larger.

Minimum-dilation graphs are not well understood yet.  As mentioned in
the introduction, it is NP-hard to decide whether there is a
$t$-spanner of a point set with at most $n-1+k$ edges, even in the
case when the spanner is a tree~\cite{chl-cmdst-07}.  However, if the
spanner is restricted to be a star then the minimum dilation graph can
be computed in polynomial time~\cite{ew-mds-05}.  Are there other
restrictions on the spanner or the point set that allow for efficient
algorithms?

Given that the general problem is NP-hard, it would also be
interesting to look for algorithms that approximate the best possible
dilation (instead of giving only a guarantee in terms of $n$ and $k$,
as we do). We are not aware of any result showing how to approximate
the minimum-dilation spanning tree with approximation factor $o(n)$.
For general \emph{unweighted} graphs, a $O(\log n)$ approximation is
possible~\cite{ep-ammss-04}.  Another result in this direction is by
Knauer and Mulzer~\cite{km-ermdt-05}, who described an algorithm that
computes a triangulation whose dilation is within a factor of $1 +
O(1/\sqrt n)$ of the optimum.

\end{document}